# Orbitally induced hierarchy of exchange interactions in zigzag antiferromagnetic state of honeycomb silver delafossite $Ag_3Co_2SbO_6$

E.A. Zvereva[1,*], M.I. Stratan[1], A.V. Ushakov[2], V.B. Nalbandyan[4], I.L. Shukaev[4], A.V. Silhanek[5], M. Abdel-Hafiez,[5,6,7], S.V. Streltsov[2,3], and A.N. Vasiliev[1,3,8]

[1]Faculty of Physics, Moscow State University, 119991 Moscow, Russia
*zvereva@mig.phys.msu.ru
[2]Institute of Metal Physics RAS, 620990 Ekaterinburg, Russia
[3]Ural Federal University, 620002 Ekaterinburg, Russia
[4]Chemistry Faculty, Southern Federal University, 344090 Rostov-on-Don, Russia
[5] Département de Physique, Université de Liége, B-4000 Sart Tilman, Belgium
[6]Institute of Physics, Goethe University Frankfurt, 60438 Frankfurt/M, Germany
[7]Faculty of Science, Physics Department, Fayoum University, 63514 Fayoum, Egypt.
[8]National University of Science and Technology "MISiS", 119049 Moscow, Russia

**Abstract**

We report the revised crystal structure, static and dynamic magnetic properties of quasi-two dimensional honeycomb-lattice silver delafossite $Ag_3Co_2SbO_6$. The magnetic susceptibility and specific heat data are consistent with the onset of antiferromagnetic long range order at low temperatures with Néel temperature $T_N \sim$ 21.2 K. In addition, the magnetization curves revealed a field-induced (spin-flop type) transition below $T_N$ in moderate magnetic fields. The GGA+U calculations show the importance of the orbital degrees of freedom, which maintain a hierarchy of exchange interaction in the system. The strongest antiferromagnetic exchange coupling was found in the shortest Co-Co pairs and is due to direct and superexchange interactions between the half-filled xz+yz orbitals pointing directly to each other. The other four out of six nearest neighbor exchanges within the cobalt hexagon are suppressed, since for these bonds active half-filled orbitals turned out to be parallel and do not overlap. The electron spin resonance (ESR) spectra reveal a Gaussian shape line attributed to $Co^{2+}$ ion in octahedral coordination with average effective g-factor g=2.3±0.1 at room temperature and shows strong divergence of ESR parameters below ~ 120 K, which imply an extended region of short-range correlations. Based on the results of magnetic and thermodynamic studies in applied fields, we propose the magnetic phase diagram for the new honeycomb-lattice delafossite.

## I. Introduction

Delafossite compounds are one of the largest families of oxides attracting a lot of attention both due to remarkable variety of their physical properties and numerous applications as catalysts, multiferroics, thermolelectric and luminescent materials, battery cathodes, transparent conductors for solar cell technologies, etc. The most studied family is $AMO_2$ oxides, where the monovalent A-site cations ($Ag^+$, $Cu^+$, $Pd^+$, or $Pt^+$) adopt linear coordination, while the M-site cation (transition metal, Al, Ga, etc) is octahedrally coordinated by oxygen and forms a triangular lattice [1]. Geometrical frustration triggered by the triangular network in presence of antiferromagnetic interactions leads to many fascinating phenomena discovered in the delafossite compounds. In particular, spectacular examples of a multiferroic behavior, the coexistence of magnetic order and ferroelectricity was found in different compounds with delafossite or closely related structures: $ACrO_2$ (A=Cu, Ag) [2-4], $AgFeO_2$ [5,6], $CuFeO_2$ [7,8] $CuFe_{1-x}M_xO_2$ (M=Al, Ga) [9-11], and $AgCrS_2$ [12-13]. Unusual spin dynamics in agreement with the $Z_2$-vortex ordering scenario was revealed for $ACrO_2$ (A=Cu, Ag, Pd) [14] and a very strong doping dependence of the magnetic properties was found in $CuMnO_2$ [15-16]. Remarkable phenomenon of dimensional crossover from anisotropic 3D (antiferromagnetic) to 2D low-energy magnetic excitations (spin-liquid like state) was found in delafossite oxide $Cu_{1-x}Ag_xCrO_2$ with an increase of x [17].

The honeycomb lattice is also a variant of structure with triangular geometry and can also be magnetically frustrated due to the next nearest neighbor interactions. As it was shown theoretically, quantum fluctuations may stabilize a quantum spin liquid (SL) phase between a state of massless Dirac fermions and an antiferromagnetically ordered insulating phase [18]. Moreover, many other exotic phases may appear in the system with the honeycomb lattice: an unusual $Z_2$ SL state, phase with quasi-molecular orbitals formed on the hexagons, incommensurate Néel order, a dimerized state with spontaneously broken rotational symmetry, a valence bond liquid and lattice nematic phases, columnar dimer order with a non-bipartite structure, etc [19-30].

In contrast to the triangular lattice delafossites mentioned above, investigations of magnetic properties for the honeycomb lattice counterparts started very recently. One can expect that simple geometrical frustration will be removed on honeycomb lattice; however, non-trivial quantum ground states might be realized due to frustrating second and third neighbor interactions. In particular, among the $Cu_3M^{2+}_2SbO_6$ series (M = Mn, Co, Ni, Cu), the magnetism of manganese compound was not characterized at all [31]. The honeycomb S=1/2 delafossite compound $Cu_5SbO_6$ ($Cu^+_3Cu^{2+}_2Sb^{5+}O_6$) was characterized as dimerized system, in which the ground state is a spin singlet [32] similarly to previously known $NaFeO_2$-derived $Cu^{2+}$ honeycomb compounds $Na_2Cu_2TeO_6$ [33-35] and $Na_3Cu_2SbO_6$ [35-39] and in spite of significant structural differences. At the same time both polytypes of $Cu_3Ni_2SbO_6$ and $Cu_3Co_2SbO_6$ compounds were found to order antiferromagnetically with transitions, at 22.3 and 18.5 K for Ni and Co variants, respectively, with Curie−Weiss fits that are in agreement with $Ni^{2+}$ and $Co^{2+}$ high spins [40, 41]. The magnetic structures of these monoclinic polytypes were determined by neutron diffraction at 4 K and described as ferromagnetic chains running along the *b*-direction which are antiferromagnetically coupled, resulting in an overall antiferromagnetic "zigzag" ordering in the honeycomb plane. The magnetic κ vector is [100] for both materials; however, the moments are aligned along the *b*-axis for the Co based compounds while for the Ni sample they are in the *ac* - plane.

In the family of silver delafossites, 19 superstructures are known now [31, 42-53], although structural characterization for 18 of them is incomplete or doubtful due to stacking faults (for details, see Table S1 of the Supplemental Material). Of these 19, seven are non-magnetic, six are $Ag_3M_2XO_6$ type with apparently honeycomb arrangement of the magnetic M cations, and the remaining six are $Ag_3MRXO_6$ type with halved content of the magnetic M cations and, thus, $MO_6$ octahedra apparently isolated from each other. Of these 12 phases with magnetic cations, no physical characterization has been performed for $Ag_3Ni_2SbO_6$ [31], $Ag_3Co_2SbO_6$ [46], $Ag_3Ni_2BiO_6$ [50], $Ag_3(NaFeSb)O_6$ [51], $Ag_3(LiMTe)O_6$ ($M^{2+}$ = Co, Ni) [52] and $Ag_3(LiMSb)O_6$ ($M^{3+}$ = Cr, Mn, Fe) [53]. The delafossite $Ag_3LiRu_2O_6$ was declared as a promising candidate for thermoelectric applications due to the combination of two-dimensionality, good conductivity and stacking disorder that is likely to correlate with low thermal conductivity [47]. Basic magnetic and transport characterization was also performed for $Ag_3LiM_2O_6$ (M=Rh, Ir) [49] and the low-spin configurations of $4d^5$ and $5d^5$ ions were supposed from analysis of magnetic response in the paramagnetic phase.

In the present work the static and dynamic magnetic properties of the quasi-2D honeycomb-lattice oxide $Ag_3Co_2SbO_6$ were investigated for the first time aiming to determine its quantum ground state, magnetic structure and to build the magnetic phase diagram.

## II. Experimental and calculation details

Ion-exchange preparation of $Ag_3Co_2SbO_6$ from $Na_3Co_2SbO_6$, its chemical and X-ray diffraction (XRD) analysis were described earlier [46]. The sample $Ag_3Zn_2SbO_6$ used here as a diamagnetic analogue for specific heat measurements, was prepared in a similar way from the lithium precursor. Completeness of the substitution was confirmed by the chemical analysis: 49,1 weight % Ag found, 48,2 % calculated. XRD powder analysis was performed with an ARL X'TRA diffractometer equipped with a solid-state Si(Li) detector using CuKα radiation. The powder patterns of the two compounds (Figs. S1 and S2 of the Supplemental Material) are almost indistinguishable. The Rietveld refinements were performed using the GSAS+EXPGUI suite [54, 55].

The magnetic measurements were performed by means of a Quantum Design SQUID – magnetometer. The temperature dependence of magnetic susceptibility was measured at the magnetic field $B = 0.1$ T in the temperature range 1.8–300 K. The isothermal magnetization curves were obtained for magnetic fields $B \leq 5$ T at $T = 1.8, 2, 2.5, 3, 5, 8, 15, 20, 30$ K after cooling the sample in zero field.

The specific heat measurements were carried out by a relaxation method using a Quantum Design PPMS system. The plate-shaped samples of $Ag_3Co_2SbO_6$ of ~0.2 mm thickness and 14.13 mg mass and $Ag_3Zn_2SbO_6$ of ~0.2 mm thickness and 9.62 mg mass were obtained by cold pressing of the polycrystalline powder. Data were collected at zero magnetic field and under applied fields up to 9 T in the temperature range 2 – 300 K.

Electron spin resonance (ESR) studies were carried out using an X-band ESR spectrometer CMS 8400 (ADANI) ($f \approx 9.4$ GHz, $B \leq 0.7$ T) equipped with a low-temperature mount, operating in the range $T = 6$–270 K. The effective g-factors of our samples have been calculated with respect to an external reference for the resonance field. We used BDPA (*a,g* - bisdiphenylene-*b*-phenylallyl) $g_{ref} = 2.00359$, as a reference material.

The band structure calculation were carried out within the GGA+U approximation [56] using full-potential linearized augmented plane wave (FP-LAPW) method as realized in the Wien2k code [57]. The exchange-correlation potential was taken in the form proposed by Perdew, Burke and Ernzerhof [58]. We chose the on-site Coulomb repulsion parameter for Co to be U = 7 eV, while Hund's rule coupling parameter ($J_H$) was taken as $J_H = 0.9$ eV [59,60]. The spin-orbit coupling for the valence shells was not taken into account in the present calculations. The Brillouin-zone (BZ) integration in the course of the self-consistency iterations was performed over 12x10x12 mesh in *k*-space. The parameter of the plane wave expansion was chosen to be $R_{MT}K_{max} = 7$, where $R_{MT}$ is the smallest muffin-tin sphere radii and $K_{max}$ - plane wave cut-off. The muffin-tin radii equal to 2.05, 1.72, 2.04 and 1.72 Bohr for Co, Sb, Ag, and O, respectively. The exchange integrals *J* were calculated by fitting the total energy difference of four magnetic configurations to the Heisenberg model written as

$$H = \sum_{ij} J_{ij} \vec{S}_i \vec{S}_j \qquad (1)$$

i.e. each pair in (1) was counted twice.

## III. Results and discussion

### *A. Structural description*

Initially, the laboratory XRD data of $Ag_3Co_2SbO_6$ and its sodium precursor were interpreted in trigonal system (space group $P3_112$) since no peak splitting indicative of symmetry lowering could be observed [46]. Nevertheless, crystal structure of $Na_3Co_2SbO_6$ was refined in the monoclinic space group C2/m [61]. Recently, using high-resolution synchrotron data, we definitely confirmed peak splitting and monoclinic symmetry for $Na_3Co_2SbO_6$, similar to $Li_3Zn_2SbO_6$ [62] and $Li_3Ni_2SbO_6$ [63]. This suggested that the structure of $Ag_3Co_2SbO_6$ may also be monoclinic. The C2/m model, although less symmetrical in the unit cell data (four lattice parameters vs. two for the trigonal cell), seems to be more symmetrical in the atomic structure, having only seven variable atomic coordinates vs. fifteen coordinates in the $P3_112$ model. Therefore, we attempted here a revision of the crystal structure of $Ag_3Co_2SbO_6$ on the C2/m model using the same experimental XRD profile.

Polyhedral view of obtained C2/m crystal structure of $Ag_3Co_2SbO_6$ is presented in Fig. 1(a). Note that both models ($P3_112$ and C2/m) have essentially identical honeycomb $[Co_2SbO_6]^{3-}$ layers (Fig. 1(b)) and dumbbell coordination of the interlayer $Ag^+$, but differ in the layer stacking mode. However, mutual positions of the two adjacent octahedral layers is the same in the two structures (Fig. 2) and the difference only becomes obvious when the third layer is added. The same holds for the two models for $Na_3M_2SbO_6$, although sodium coordination is octahedral rather than linear. This enables formation of multiple stacking faults, especially characteristic for silver compounds due to (i) larger interlayer distances (e.g., 5.36 Å for $Na_3Co_2SbO_6$ and 6.22 Å for $Ag_3Co_2SbO_6$) and (ii) layer gliding during the ion exchange driving cation coordination from octahedral to linear.

Due to multiple stacking faults, the monoclinic phases effectively imitate trigonal symmetry, and their superlattice reflections responsible for $M^{2+}/Sb^{5+}$ ordering are anomalously diffuse and weak for both $Co^{2+}$ and $Zn^{2+}$ compounds (see Figs. S1 and S2 of the Supplemental Material). This is a typical feature of other compounds of this class listed in Table S1 of the Supplemental Material. This feature hinders their structural investigation. None of the 18 silver delafossite superstructures listed there, prepared by ion exchange, could receive adequate structural description. Note that $Ag_3Zn_2SbO_6$ was prepared earlier from the same precursor but no superlattice effects were observed although the authors anticipated them [31] and the starting $Li_3Zn_2SbO_6$ showed unambiguous Zn/Sb ordering [62].

Results of our structural refinements for both $Co^{2+}$ and $Zn^{2+}$ compounds are shown in Figs. S1 and S2 and listed in Tables S2, S3 and S4 of the Supplemental Material and in Crystallographic information files (cif). Reasonable agreement between experimental and calculated profiles could only be obtained with considerable $M^{2+}/Sb^{5+}$ mixing on octahedral positions. However, we suppose that this substitution is only fictitious due to stacking faults, and each individual layer is essentially ordered as in many other similar superstructures [47, 48, 64-67]. Note that average metal-oxygen distances for nominal ($Sb_{0.46}Co_{0.54}$) and ($Co_{0.73}Sb_{0.27}$) sites, 1.88 and 2.21 Å, respectively, differ significantly and do not support mixing, since ionic radii sums for $Sb^{5+}$-O and $Co^{2+}$-O bonds are 1.98 and 2.12 Å, respectively. The Co and Zn compounds are strictly isostructural and even have very similar degree of the apparent $M^{2+}/Sb^{5+}$ substitution. The polyhedral view of the $Ag_3Co_2SbO_6$ crystal structure, ignoring this mixing, is shown in Figs. 1-2. The refined silver site occupancies in $Ag_3Co_2SbO_6$ are slightly lower than unity in accordance with the analytical data: 47.3 % Ag found [46], 49.3% calculated from the ideal formula.

### B. Magnetic susceptibility and magnetization

The temperature dependence of magnetic susceptibility $\chi(T)$ implies an antiferromagnetic behavior. It was established, that upon a decrease of the temperature the $\chi(T)$ dependence at the field $B = 0.1$ T passes through a sharp maximum at $T_{max} \sim 25.2$ K and then drops (Fig. 3(a)). Such a behavior indicates an onset of antiferromagnetic long-range ordering in the material at low temperature. The $\chi(T)$ dependence recorded upon cooling the sample in zero magnetic field (ZFC) mode and in magnetic field (FC) one shows practically no divergence down to ~7 K. In applied magnetic fields the position of the $T_{max}$ is slightly shifted to the low temperature side (Fig. 3(b)).

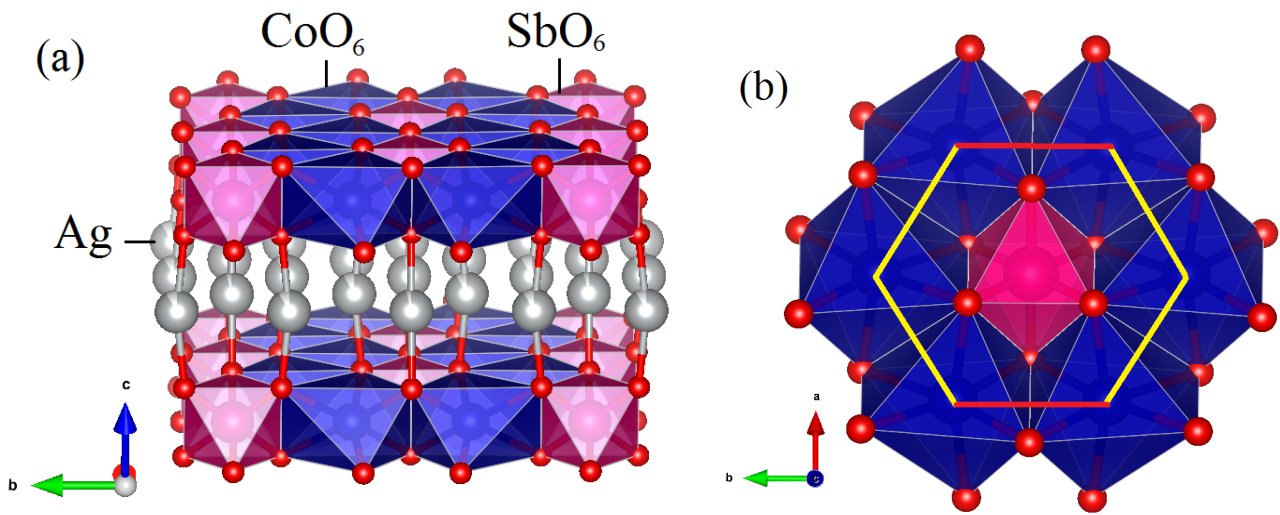

Fig. 1. Polyhedral view of a layered C2/m crystal structure of $Ag_3Co_2SbO_6$ (a) and its fragment within the magneto-active layers, assuming complete cation ordering (ignoring fictitious Co/Sb mixing) (b). $CoO_6$ and $SbO_6$ octahedra are shown by blue and pink color, respectively. Silver and oxygen ions are gray and red spheres, respectively. Long and short Co-Co distances are denoted by yellow and red lines, respectively.

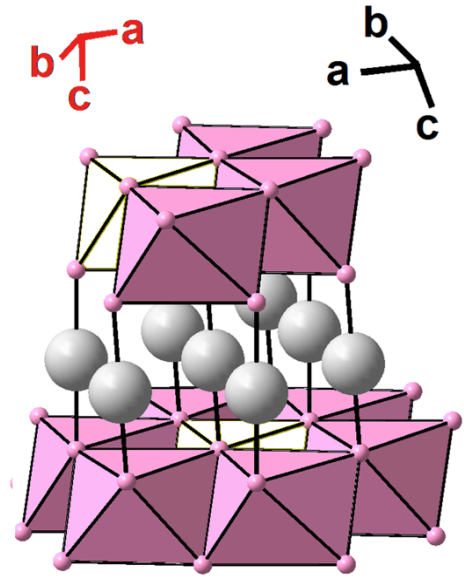

Fig. 2. Two adjacent layers in $Ag_3Co_2SbO_6$, similar arrangement of the two structural models: $P3_112$ and $C2/m$, with corresponding coordinate systems on left and right, respectively. $SbO_6$ and $CoO_6$ octahedra are white and pink, respectively, and silver ions are shown as gray balls.

At high temperatures, the magnetic susceptibility nicely follows the Curie-Weiss law with addition of a temperature-independent term $\chi_0$. The best fit of the experimental data in the range 200-300 K yields $\chi_0$ = 0.012(3) emu/mol, $C$ = 6.2(5) emu/mol K and the Weiss temperature $\Theta$ = -9(1) K indicating a predominance of the antiferromagnetic interaction in the compound. The effective magnetic moment $\mu_{eff}$ estimated from the Curie constant $\mu_{eff}$ is about 7 $\mu_B$/f.u., which is slightly higher than one expected from theoretical estimations for system of two $Co^{2+}$ ions obtained by using g=2.3±0.1 experimentally determined from the ESR spectra at room temperature (see below).

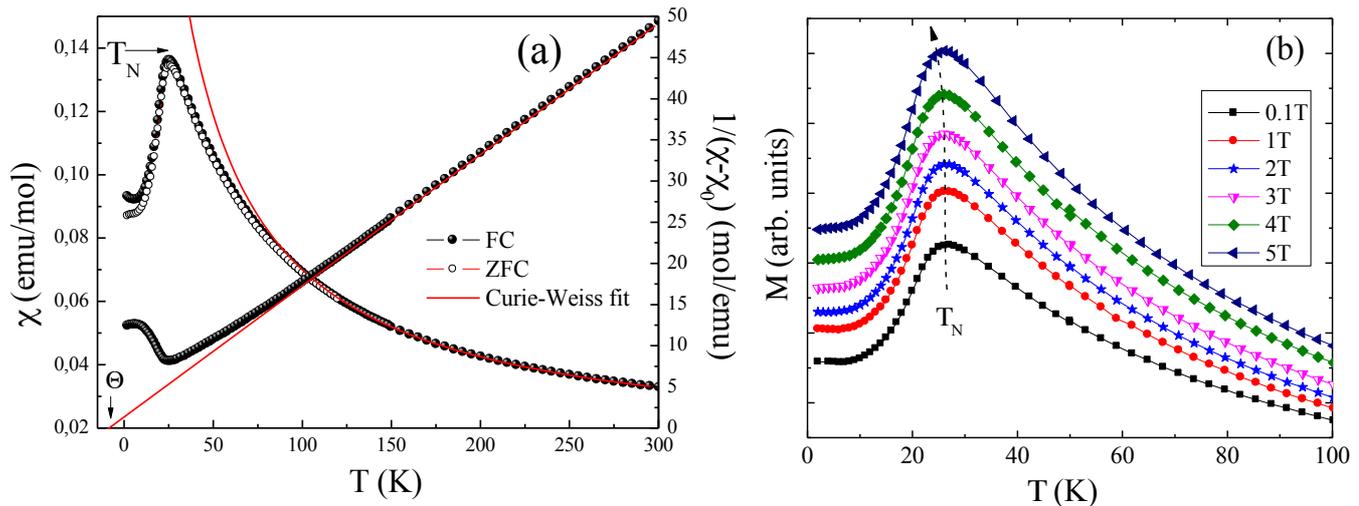

Fig. 3. (a) Temperature dependence of the magnetic susceptibility recorded in ZFC (open circles) and FC (filled circles) regimes at $B$ = 0.1 T for $Ag_3Co_2SbO_6$. The solid curve represents an approximation in accordance with the Curie-Weiss law. (b) $M(T)$ curves at various external magnetic fields. The dashed arrow indicates the shift of $T_N$ upon variation of the magnetic field.

The full $M$(B) isotherm at $T = 1.8$ K for $Ag_3Co_2SbO_6$ shows no hysteresis and no saturation in magnetic fields up to 5 T (Fig. 4). It is worth noting, that within this range of the applied magnetic fields, the magnetic moment is still below the theoretically expected saturation magnetic moment for two $Co^{2+}$ ions per formula unit assuming either high-spin (S=3/2) or low-spin state (S=1/2): $M_s = 2gS\mu_B$. One can see, however, that the magnetization curve has a slight upward curvature suggesting the presence of a magnetic field induced spin-flop type transition. The value of critical field has been estimated from the maximum in the derivative d$M$/d$B$($B$) curve as high as $B_C \sim 1.9$ T (see an inset in Fig. 4). With increasing temperature, the amplitude of this anomaly decreases rapidly, and the feature becomes undetectable above $T_N$ (Fig. S3). It is worth to mention that the spin-reorientation-type transitions have been reported recently for several related antimonates: for example, $Na_3Co_2SbO_6$ [61], $Li_3Ni_2SbO_6$ [63], $Na_3Ni_2SbO_6$ [76], $Li_4FeSbO_6$ [67].

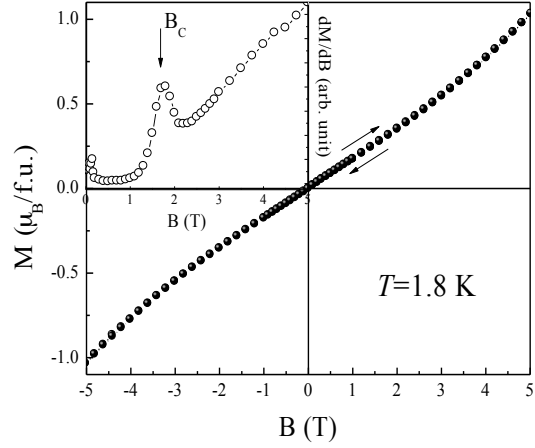

Fig. 4. The full magnetization isotherm for $Ag_3Co_2SbO_6$ and its first derivative d$M$/d$B$ (on inset) at $T$=1.8 K.

*C. Specific heat*

The specific heat data $C(T)$ for $Ag_3Co_2SbO_6$ at zero magnetic field are in good agreement with the temperature dependence of magnetic susceptibility in weak magnetic fields, and demonstrate a distinct λ-shaped anomaly, which is characteristic of a three-dimensional (3D) magnetic order (Fig. 5). Note, however, that the absolute value of the Néel temperature $T_N \sim 21.2$ K, deduced from $C(T)$ data at $B = 0$ T is slightly lower than $T_{max} \sim 25.2$ K on $\chi(T)$ dependence at $B = 0.1$ T (Fig. 5), whereas it correlates with a maximum on the magnetic susceptibility derivative $\partial\chi/\partial T(T)$. Indeed, as it has been shown by Fisher [68,69] the temperature dependence of the specific heat $C(T)$ for the antiferrimagnets with short-range interactions should follow the derivative of the magnetic susceptibility in accordance with:

$$C(T) = A\left(\frac{\partial}{\partial T}\right)\left[T\chi_{\parallel}(T)\right] \quad (2)$$

where the constant A depends weakly on temperature. In accordance with Eq. (2), the λ-type-anomalies observed in $C(T)$ dependence at the antiferromagnetic transition temperature are defined by an infinite positive gradient on the curve $\chi_{\parallel}(T)$ at $T_N$, while a maximum $\chi_{\parallel}(T)$ lies usually slightly above the ordering temperature. Thus, the anomaly in the specific heat should correspond to the similar anomaly in $\partial\chi_{\parallel}/\partial T(T)$ [70].

In order to analyze the nature of the magnetic phase transition and to evaluate the corresponding contribution to the specific heat capacity and entropy, specific heat was also measured for the strictly isostructural diamagnetic material $Ag_3Zn_2SbO_6$. The specific heat data for both magnetic and diamagnetic samples in the $T$-range 2-300 K are shown in Fig. 6. The Dulong-Petit value reaches 3Rz = 299 J/mol K, with the number of atoms per formula unit z = 12. We observe a specific heat jump of $\Delta C_p \sim 14.4$ J/mol K at $T_N$, which is a noticeably smaller than the value expected from the mean-field theory for the antiferromagnetic ordering of two $Co^{2+}$ ions system assuming all ions to be in the high-spin (S=3/2) state [71]: $\Delta C_p = 5R \dfrac{2S(S+1)}{S^2 + (S+1)^2} \approx 36.7$ J/mol×K where $R$ being the gas constant $R$=8.31 J/mol K. Note, that such a reduction of $\Delta C_p$ may indicate the presence of short-range magnetic correlations at higher temperature in $Ag_3Co_2SbO_6$. In applied magnetic fields, the $T_N$ - anomaly is slightly rounded and markedly shifts to the lower temperature (see inset in Fig. 6).

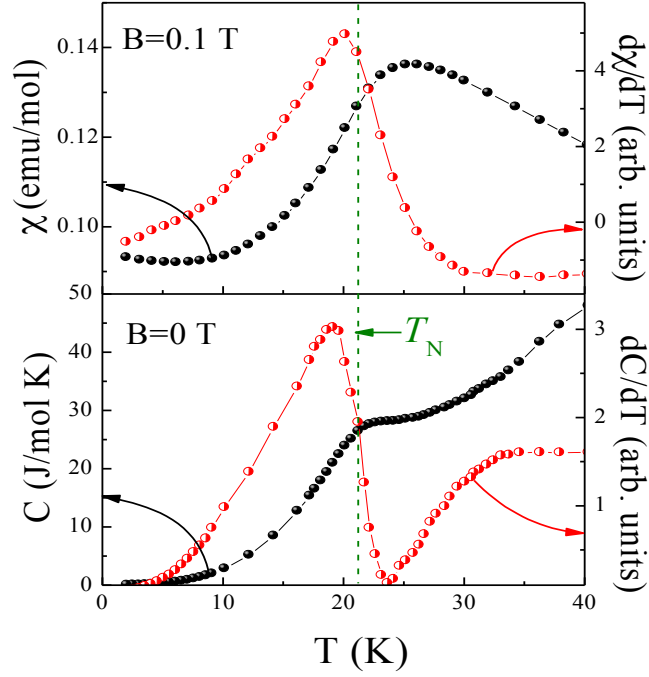

Fig. 5. Temperature dependence of the magnetic susceptibility at $B$ = 0.1 T, and the specific heat at $B$ = 0 T and their derivatives for $Ag_3Co_2SbO_6$.

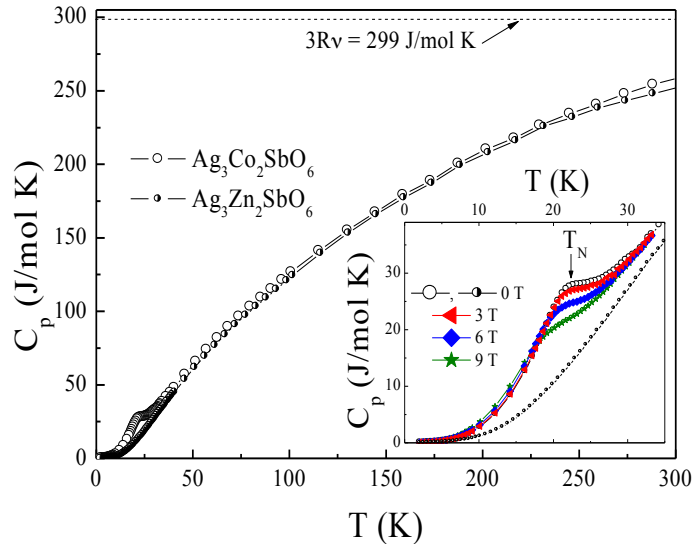

Fig. 6. Temperature dependence of the specific heat in $Ag_3Co_2SbO_6$ (black open circles) and the isostructural diamagnetic compound $Ag_3Zn_2SbO_6$ (black half-filled circles) in zero magnetic field. Inset: enlarged low temperature part highlights the onset of antiferromagnetic spin ordering and shift of the $T_N$ in magnetic fields.

For quantitative estimations we assume that the specific heat of the isostructural compound $Ag_3Zn_2SbO_6$ provides a proper estimation for the pure lattice contribution to specific heat. In the frame of Debye model the phonon specific heat is described by the function [71]:

$$C_{ph} = 9R\left(\frac{T}{\theta_D}\right)^3 \int_0^{T/\theta_D} \frac{e^x x^4}{(e^x - 1)^2} dx \qquad (4)$$

where $x = \hbar\omega/kT$, $\theta_D = \hbar\omega_{max}/k$ – Debye temperature, $\omega_{max}$ - maximum frequency of the phonon spectrum, $k$ – Boltzmann constant. The value of Debye temperature $\theta_D$ estimated from approximation of C(T) according to this $T^3$ – law in low temperature range for the diamagnetic compound $Ag_3Zn_2SbO_6$ was found to be about ~ 285 ± 5 K. Normalization of the Debye temperatures has been made taking into account the difference between the molar masses for Zn – Co atoms in the $Ag_3Co_2SbO_6$ compound resulting in $\theta_D$ ~ 286 ± 5 K for $Ag_3Co_2SbO_6$.

The magnetic contribution to the specific heat was determined by subtracting the lattice contribution using the data for the isostructural non-magnetic analogue (Fig. 7). We examine the $C_m(T)$ below $T_N$ in terms of the spin-wave (SW) approach assuming the limiting low-temperature behavior of the magnetic specific heat should follow $C_m \propto T^{d/n}$ - power law for magnons [66], where $d$ stands for the dimensionality of the magnetic lattice and $n$ is defined as the exponent in the dispersion relation $\omega \sim \kappa^n$. For antiferromagnetic (AFM) and ferromagnetic (FM) magnons $n = 1$ and $n = 2$, respectively. The least square fitting of the data below $T_N$ (upper panel in Fig. 7) has given with good accuracy $d = 3$ and $n = 0.9$ values, that corroborates the picture of 3D AFM magnons at the low temperatures.

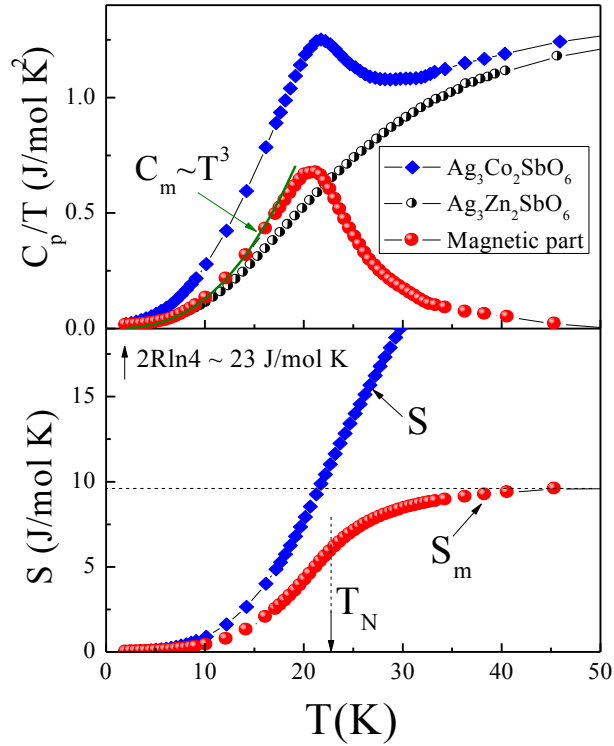

Fig. 7. Magnetic specific heat (red filled circles on upper panel) and magnetic entropy (red filled circles on lower panel) in comparison with the specific heat data for $Ag_3Co_2SbO_6$ (blue diamonds) and non-magnetic analogue $Ag_3Zn_2SbO_6$ (black half-filled circles) at $B=0$ T. Solid curve indicates the spin wave contribution estimated in accordance with $C_m \propto T^{d/n}$ - power law for magnons.

The entropy change has been calculated using the equation: $\Delta S_m(T) = \int_0^T \frac{C_m(T)}{T} dT$ (lower panel in Fig. 7). One can see that the magnetic entropy $\Delta S_m$ saturates at about 40 K, reaching approximately 9.6 J/(mol K). This value is essentially lower than the magnetic entropy change expected from the mean-field theory for system of two cobalt magnetic ions with S=3/2: $\Delta S_m(T) = 2R\ln(2S+1) \approx 23$ J/(mol K). One should note that the magnetic entropy released below $T_N$ removes only about 26% of the saturation value. This indicates the presence of appreciable short-range correlations far above $T_N$, which is usually characteristic feature for materials with lower magnetic dimensionality [70].

### D. ESR spectra

The ESR data are in satisfactory agreement with static magnetization ones. ESR spectra in the paramagnetic phase ($T > T_N$) show a single broad Gaussian shape line (Fig. 8(a)) ascribable to $Co^{2+}$ ions in octahedral coordination. The shape of the spectrum changes noticeably with decreasing temperature. One can see that below ~ 100 K visible broadening and distortion of the absorption line occurs. The intensity of the ESR signal increases with decrease of the temperature in the paramagnetic phase, passes through a maximum in the vicinity of the Néel temperature, then decreases and eventually vanishes at low temperatures. This low-$T$ signal fading could be indicative of an opening of an energy gap for resonance excitations, e.g., due to the establishment of AFM order.

Since the line is relatively broad (only one order less than the resonance field in the present compound), two circular components of the exciting linearly polarized microwave field have to be taken into account. Therefore, for analysis, the ESR signals on both sides of $B = 0$ have to be included into the fit formula, which has been taken in conventional form:

$$\frac{dP}{dB} \propto \frac{d}{dB}\left[\exp\left(\frac{(-\ln 2)(B-B_r)^2}{\Delta B^2}\right) + \exp\left(\frac{(-\ln 2)(B+B_r)^2}{\Delta B^2}\right)\right] \quad (3)$$

where $P$ is the power absorbed in the ESR experiment, $B$ – magnetic field, $B_r$ – resonance field, $\Delta B$ – the linewidth. Results of ESR lineshape fitting in accordance with Eq. (3) are shown by solid lines in Fig. 8(a). Apparently, the fitted curves are in reasonable agreement with the experimental data.

It was established, that at high temperatures ($T > 120$ K) the absorption is characterized by almost temperature independent values of the effective g-factor g=2.3±0.1 and the linewidth $\Delta B \approx 100$ mT (Fig. 8(b)). The theory for octahedral field predicts that the EPR spectrum from high-spin $Co^{2+}$ may be observed mainly at low temperatures because of spin-lattice relaxation time is extremely short for octahedral coordination of $Co^{2+}$, while at high temperature spectra are usually undetectable to its strong broadening. The expected effective g-factor for high-spin $Co^{2+}$ is 4.3 in ideal oxygen octahedral environment [67,68] or takes lower value in distorted octahedral coordination. Below certain characteristic temperature $T_{SRO}$ both g-factor and the linewidth increase progressively upon lowering of the temperature and reveal clear anomalies in the vicinity of Néel temperature. A pronounced broadening and the visible shift of the resonant field to lower magnetic fields indicate a large pre-ordering magnetic effect in $Ag_3Co_2SbO_6$ in the $T$-range 25 – 120 K, most probably due to strong short-range magnetic fluctuations essentially higher than ordering temperature, those are characteristic of the low-dimension magnetic systems. The presence of such fluctuations is evident also from the static susceptibility data showing visible deviation form the Curie-Weiss law at the same temperature (Fig. 3(a)).

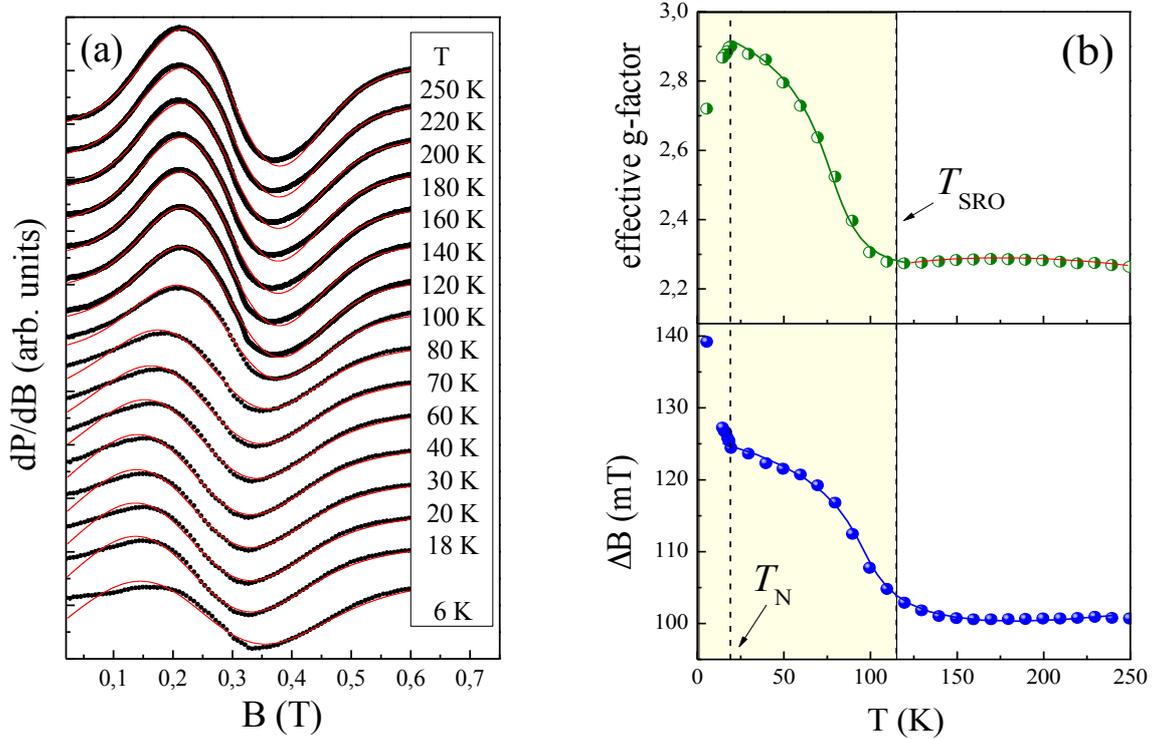

Fig. 8. (a) Temperature evolution of the first derivative ESR absorption line for $Ag_3Co_2SbO_6$: black points – experimental data, lines – fitting in accordance with Gaussian profile (Eq. 3). (b) Temperature dependence of the effective g-factor and the ESR linewidth for $Ag_3Co_2SbO_6$.

*E. Band structure calculations and analysis of the orbital structure.*

The total and partial density of states of $Ag_3Co_2SbO_6$ obtained in the GGA+U calculations are presented in Fig. 9. One may see that for given values of $U$ and $J_H$ parameters this compound is insulating with the band gap ~1 eV. The top of the valence band is defined by O-p, while the bottom of the conduction band by the Ag-s,p,d and Co-d states. Spin moment on Co was found to be $2.7\mu_B$, which is close to the ionic value for $Co^{2+}$, but reduced due to the hybridization, which is typical for transition metal compounds.

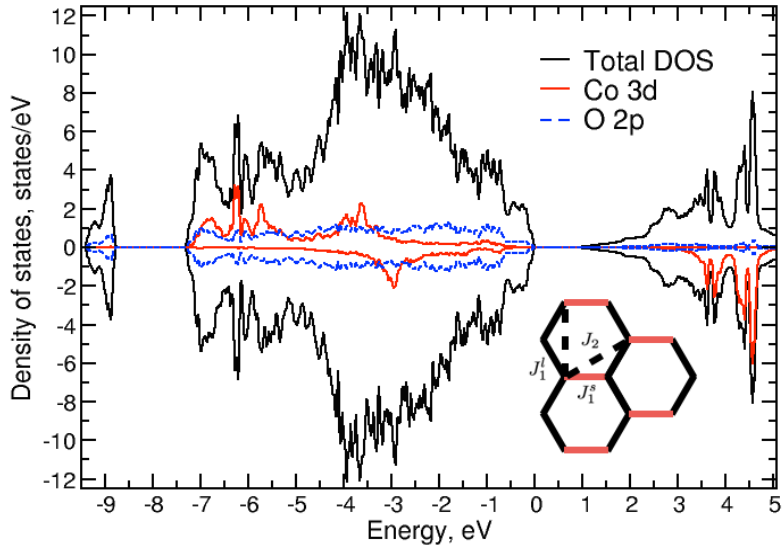

Fig. 9. Total and partial density of states for $Ag_3Co_2SbO_6$ as obtained in the GGA+U calculation for the AFM zigzag order. Total density of states is normalized per formula unit, while partial – per ion. Upper (lower) panel corresponds to spin majority (minority). The Fermi energy is set to zero. The inset shows how the exchange parameters are defined: $J_1^l$ is between long (black) nearest neighbors, $J_1^s$ is between short (red) nearest neighbors Co ions, $J_2$ is the average exchange between next nearest neighbors in the hexagonal plane.

We calculated four magnetic configurations, which are shown in Fig. 1 of Ref. 76 in order to extract three exchange integrals: $J_1^s$, $J_1^l$, and $J_2$ in the $ab$ plane. First two exchange parameters, $J_1^s$ and $J_1^l$, are between nearest neighbors in the honeycomb plane; $J_1^s$ is along the shortest Co-Co pathways (there are two such pairs in each hexagon), while $J_1^l$ characterizes magnetic coupling for long Co-Co pairs (4 bonds in each hexagon). $J_2$ is the average exchange between next nearest neighbors in the hexagonal plane.

It was found that similarly to related honeycomb lattice antimonates $A_3Ni_2SbO_6$ (A = Li, Na) [76] and the lowest total energy corresponds to the AFM zigzag state. It is important to mention that this type ordering has been recently evidenced for honeycomb-lattice delafossites $Cu_3Ni_2SbO_6$ and $Cu_3Co_2SbO_6$ as refined experimentally from low temperature neutron diffraction studies [40,41]. In $Ag_3Co_2SbO_6$ the exchange parameters were calculated to be $J_1^s$ =28 K (AFM), $J_1^l$ =-2 K (FM), and $J_2$=3 K (AFM). This is rather intriguing, that two nearest neighbor exchanges, $J_1^s$ and $J_1^l$, are so different. Indeed, the difference in the Co-Co lengths is not so large: $\delta r$ ~0.016 A. If one assumes that $J_1 \sim 2t^2/U$ and the $d$-$d$ hopping $t$ follows conventional Harrison's dependence $t_d \sim 1/r^5$ [77] ($r$ here is the distance between Co), than this change in $r$ would result only ~ 5% difference in the exchange. An account of the superexchange contribution and $r$-dependence of $p$-$d$ hopping does not improve the situation considerably, so that the origin of the strong difference in nearest neighbor exchanges has to be due to some other mechanism.

In order to find this mechanism we analyzed the orbital structure. As it was mentioned above Co ions have octahedral surrounding in $Ag_3Co_2SbO_6$. Moreover, these octahedra turned out to be compressed, which results in a particular splitting of the Co $3d$ shell. If one would chose local coordinate system as shown in Fig. 10 (i.e. all axes are directed to the oxygens; $z$-axis is along the shortest Co-O bonds), than $t_{2g}$ subband is split on lower lying in energy $xy$ and higher $zx/yz$ subbands. In the case of $3d^7$ configuration ($Co^{2+}$) this is the $zx/yz$ subband, which turns out to be 3/4-filled and hence "magnetically active". This is the feature of the $Ag_3Co_2SbO_6$ crystal structure, that the shortest Co-O bonds are directed very differently in pairs of the $CoO_6$ octahedra forming short and long Co-Co nearest neighbor bonds. This leads to a specific orbital order, which in turn results in very different exchange parameters for these bonds. We used Linearized muffin-tin orbitals (LMTO) method [78] to plot the single half-filled $t_{2g}$ orbital in the LDA+U calculation. This method is based on the atomic-like wave functions, which makes such a procedure straightforward (the same method was used previously, e.g. in $CaCrO_3$ [79] and $NaMn_7O_{12}$ [80]). The resulting orbital order is shown in Fig. 10. One may see that the single half-filled $t_{2g}$ orbitals for two Co ions forming short (red in Figs. 1 and 10) Co-Co bonds are directed to each other (these are the $xz+yz$ orbitals in the coordinate system with $z$-axis directed along short, (blue) Co-O bond and $x$ and $y$ axes pointing to oxygens forming common edge). Such an orbital pattern will result in AFM exchange via both direct and superexchange mechanisms. In contrast, if one considers long Co-Co bonds than it turns out that these half-filled $xz+yz$ orbitals do not overlap with each other, i.e. direct AFM exchange via long Co-Co bonds is suppressed by this orbital order. Moreover, the superexchange via the same $p$-orbital of oxygen is also impossible due to signs of the $p$- and $d$-wave functions. The superexchange via different $p$-obitals is possible, but it is usually small and FM [81]. Since $xy$-orbital on each Co is fully occupied the $t_{2g}$-$e_g$ exchange is vanishingly small. The superexchange between $e_g$ orbitals is also not operative in the common edge geometry [81,82].

Thus, the analysis performed above shows that the orbital order realizing in $Ag_3Co_2SbO_6$ blocks AFM $t_{2g}$-$t_{2g}$ exchange between part of the nearest neighbor Co ions. The situation here reminds $NaTiSi_2O_6$, where features of the crystal structure strongly affect the nearest neighbor exchange in such a way that half-filled $d$-orbitals in part of the metal-metal pairs turn out to be "parallel" to each other and do not contribute to the exchange coupling [81,83,84]. The analysis of orbital structure in $Ag_3Co_2SbO_6$ fully agrees with estimations of the exchange integrals obtained with the use of the total energy calculations according to which there is only one strong exchange coupling in this system and this is the exchange corresponding to the shortest Co-Co bonds.

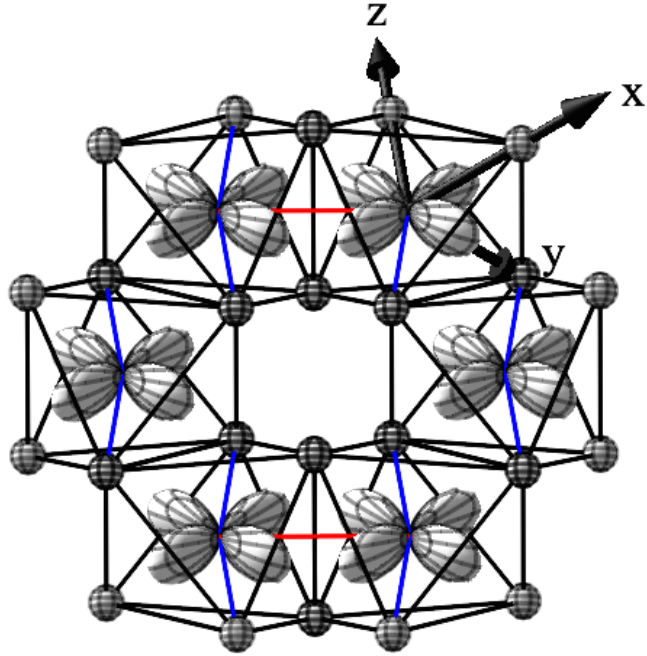

Fig. 10. (Color online) The orbital order as obtained in the LDA+U calculations (in the LMTO). The single half-filled $t_{2g}$ orbital of Co is shown. O ions are shown as balls, short Co-Co bonds are painted in red, short Co-O bonds in blue. Local coordinate system in one of the $CoO_6$ octahedra is shown.

### F. Magnetic phase diagram

Summarizing the data of thermodynamic studies, performed in the present work, the magnetic phase diagram for the new honeycomb lattice dellafossite $Ag_3Co_2SbO_6$ was constructed (Fig. 11). In zero magnetic field the paramagnetic phase is realized at temperatures higher than 21.2 K, while applying the magnetic field slightly shifts this phase boundary toward the lower temperatures, which is typical for antiferromagnets. It is also evident from the *B-T* diagram that one more magnetic phase (II) is induced by applying a magnetic field below $T_N$ and probably associated with different mutual orientations of neighboring spins in honeycomb lattice. The quantum ground state of $Ag_3Co_2SbO_6$ was determined as *zigzag* antiferromagnetic state (I) exists below ~ 2.7 T.

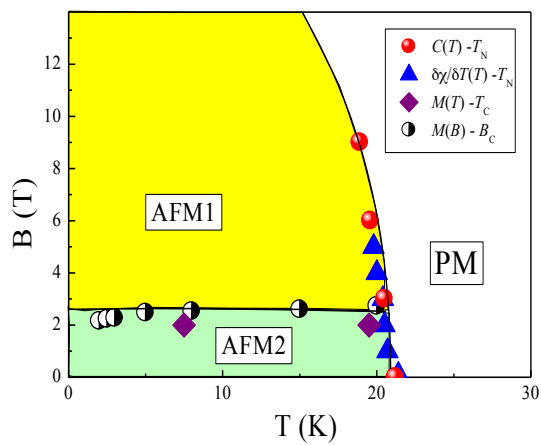

Fig. 11. Magnetic phase diagram for antimonate $Ag_3Co_2SbO_6$ as defined from thermodynamic (magnetization d$M$/d$T$ and d$M$/d$B$ and specific heat $C(T)$) measurements.

### IV. Conclusion

In conclusion, we have investigated the magnetic properties of a new quasi 2D honeycomb-lattice layered silver delafossite $Ag_3Co_2SbO_6$. The static magnetic susceptibility and specific heat data show the onset of antiferromagnetic order at $T_N \sim 21.2$ K. The entropy release occurs mainly at temperatures higher than $T_N$, indicating the presence of appreciable short-range correlations in the compound. The high-temperature magnetic susceptibility data exhibits Curie-Weiss behavior with a Weiss temperature $\Theta \sim -9$ K that indicates a predominance of the antiferromagnetic coupling. ESR spectra in the paramagnetic phase show a single Gaussian shape line attributed to $Co^{2+}$ ions in octahedral coordination characterized by the isotropic effective g-factor $g = 2.3\pm0.1$. However, the distortion, broadening of the ESR absorption line and shift of the resonant field were found to take place below $\sim 120$ K, which imply complex dynamic properties and an extended region of short-range order correlations in this quasi 2D compound. The theoretical calculations show that the strongest exchange interaction is between the nearest neighbors, but this nearest neighbors exchange is essentially affected by a specific orbital structure, which suppresses antiferromagnetic exchange in four out of six (nearest neighbor) Co-Co pathways within the cobalt hexagon. It was also found that both ferromagnetic and antiferromagnetic intraplane exchange interactions are present on the honeycomb $Co_2SbO_6$ layers and the most favorable spin configuration model is weakly ferromagnetic chains, which are coupled via strong antiferromagnetic exchange resulting in overall *zigzag* AFM order in $Ag_3Co_2SbO_6$.

## Acknowledgments

E.A.Z., V.B.N. and I.L.S. appreciate support from the Russian Foundation for Basic Research (grants14-02-00245 and 11-03-01101). E.A.Z., S.V.S. and A.V.U. appreciate support from the Russian Foundation for Basic Research (grant 13-02-00374), S.V.S. and A.N.V. also acknowledge support from the FASO (theme "Electron" No. 01201463326) and Government of the Russian Federation, contract № 02.A03.21.0006. This work was supported in part from the Ministry of Education and Science of the Russian Federation in the framework of Increase Competitiveness Program of NUST «MISiS» (№ К2-2015-075). The work in Germany by M.A. was supported by the DFG Eigene Stelle MO 3014/1-1. The work of M.A. and A.V.S. was supported by the FNRS projects, credit de demarrage U.Lg."

Supplementary material for the paper

# Orbitally induced hierarchy of exchange interactions in zigzag antiferromagnetic state of honeycomb silver delafossite $Ag_3Co_2SbO_6$

E.A. Zvereva[1,*], M.I. Stratan[1], A.V. Ushakov[2], V.B. Nalbandyan[4], I.L. Shukaev[4], A.V. Silhanek[5], M. Abdel-Hafiez,[5,6,7] S.V. Streltsov[2,3], and A.N. Vasiliev[1,3,8]

[1]Faculty of Physics, Moscow State University, 119991 Moscow, Russia
*zvereva@mig.phys.msu.ru
[2]Institute of Metal Physics, S. Kovalevskoy St. 18, 620990 Ekaterinburg, Russia
[3]Ural Federal University, 620002 Ekaterinburg, Russia
[4]Chemistry Faculty, Southern Federal University, 344090 Rostov-on-Don, Russia
[5] Département de Physique, Université de Liége, B-4000 Sart Tilman, Belgium
[6]Institute of Physics, Goethe University Frankfurt, 60438 Frankfurt/M, Germany
[7]Faculty of Science, Physics Department, Fayoum University, 63514 Fayoum, Egypt.
[8]National University of Science and Technology "MISiS", 119049 Moscow, Russia

Contents

Table S1. Known $Ag_3M_2RO_6$ and $Ag_3MRXO_6$ delafossite-related mixed oxides

Only $Ag_2SnO_3$ [37, 38] has been prepared by direct synthesis and characterized by single-crystal diffraction. All other compounds have been prepared by ion exchange from the sodium or lithium counterparts; they exhibit very similar powder diffraction patterns typical of the rhombohedral $AgFeO_2$ type; almost all show one or two weak and diffuse superstructure reflection; however, interpretations of these patterns are different.

| Composition | Comments |
| --- | --- |
| $Ag_3(AgSn_2)O_6$ [42, 43] | The basic structure is double-layered $P6_322$ honeycomb type, but there is an incommensurate modulation with a = 29.22 Å. |
| $Ag_3LiTi_2O_6$ [44] | Powder pattern indexed as a trigonal superstructure. No structure refinement. |
| $Ag_3M_2SbO_6$ (M = Ni, Zn) [31] | No superlattice reflections found. The structures were refined within disordered $Ag(M_{2/3}Sb_{1/3})O_2$ $R\bar{3}m$ model, although the authors assumed high degree of local order. |
| $Ag_3LiM_2O_6$ (M = Ti, Sn) [45] | No indexing; disordered $Ag(Li_{1/3}M_{2/3})O_2$ formulas were used although one or two superlattice reflections were visible. |
| $Ag_3Co_2SbO_6$ [46] | The superstructure was refined within the trigonal $P3_112$ model. However, accuracy was low, and monoclinic symmetry could not be excluded as discussed in the present paper. |
| $Ag_3LiRu_2O_6$ [47] | Paradoxically, superstructure was refined omitting the unique superstructure reflection; monoclinic $C2/m$ model was used although no splitting of reflections from the rhombohedral subcell was detected. It was assumed that each layer was ordered but their stacking was disordered. |
| $Ag_3LiRu_2O_6$ [48] | No superlattice reflections found. The structure was refined within disordered $Ag(Li_{1/3}Ru_{2/3})O_2$ $R\bar{3}m$ model. Lattice parameters differ significantly (by 2.0-2.4%) from the subcell parameters in the preceding work [42]. |
| $Ag_3LiM_2O_6$ (M = Rh, Ir) [49] | Superlattice reflections were found in both X-ray and electron diffraction. Each layer is assumed to be fully ordered. However, the structures were refined within disordered $Ag(Li_{1/3}M_{2/3})O_2$ $R\bar{3}m$ model. |
| $Ag_3Ni_2BiO_6$ [50] | Indexed as a trigonal superstructure ($P3_112$). No structure refinement. |
| $Ag_3NaFeSbO_6$ [51] | Indexed as a trigonal superstructure ($P3_112$). Structural model was depicted but not refined. |
| $Ag_3LiMTeO_6$ (M = Co, Ni, Zn) [52], $Ag_3LiMnSbO_6$ [53] | Superlattice reflections are visible but not discussed. No indexing, no lattice parameters. |
| $Ag_3LiMSbO_6$ (M = Al, Cr, Fe, Ga) [53] | Indexed as a trigonal superstructure ($P3_112$). No structure refinement. |

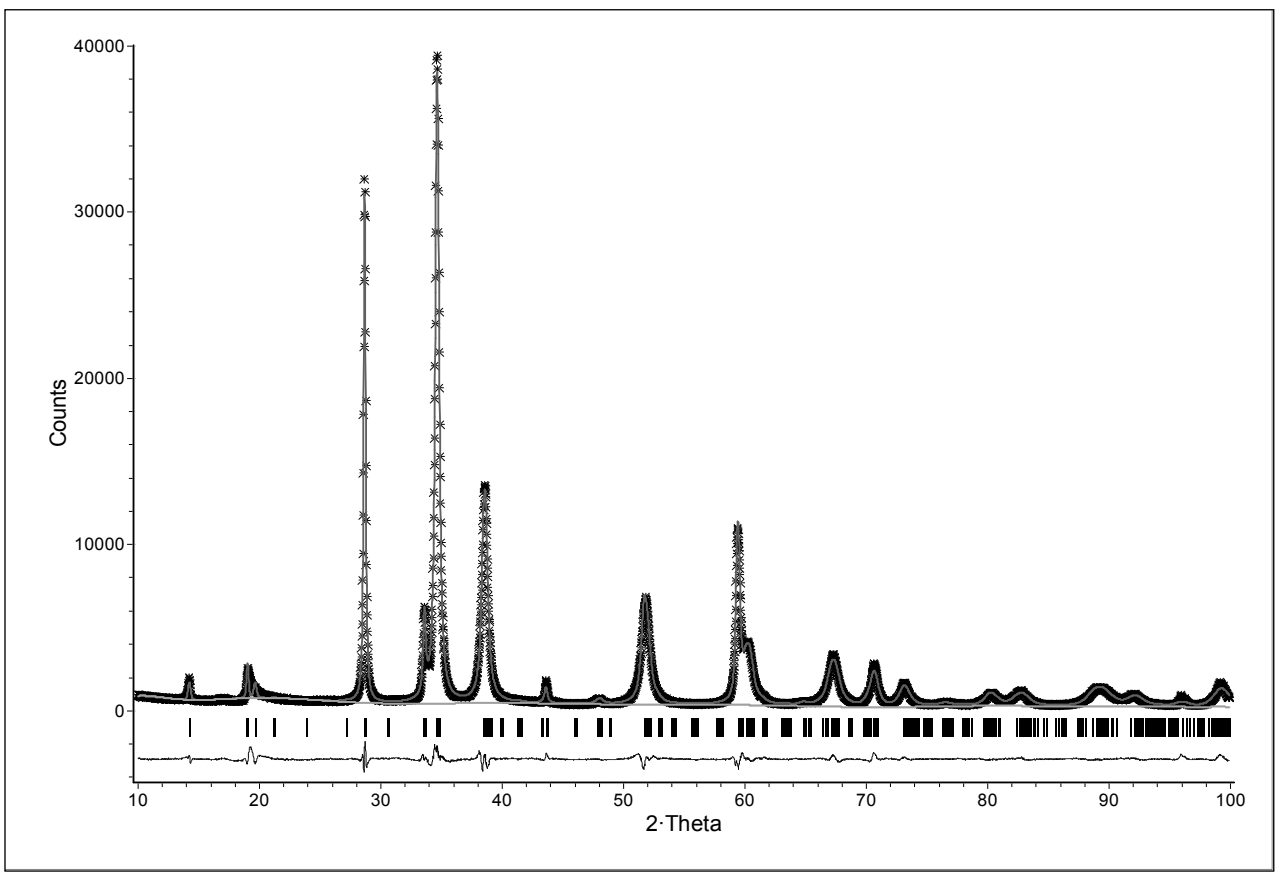

Fig. S1. XRD profile of $Ag_3Co_2SbO_6$.

Stars – experimental data, dark gray line – calculated pattern, light gray line – background, vertical bars – Bragg ticks, thin black line below – difference.

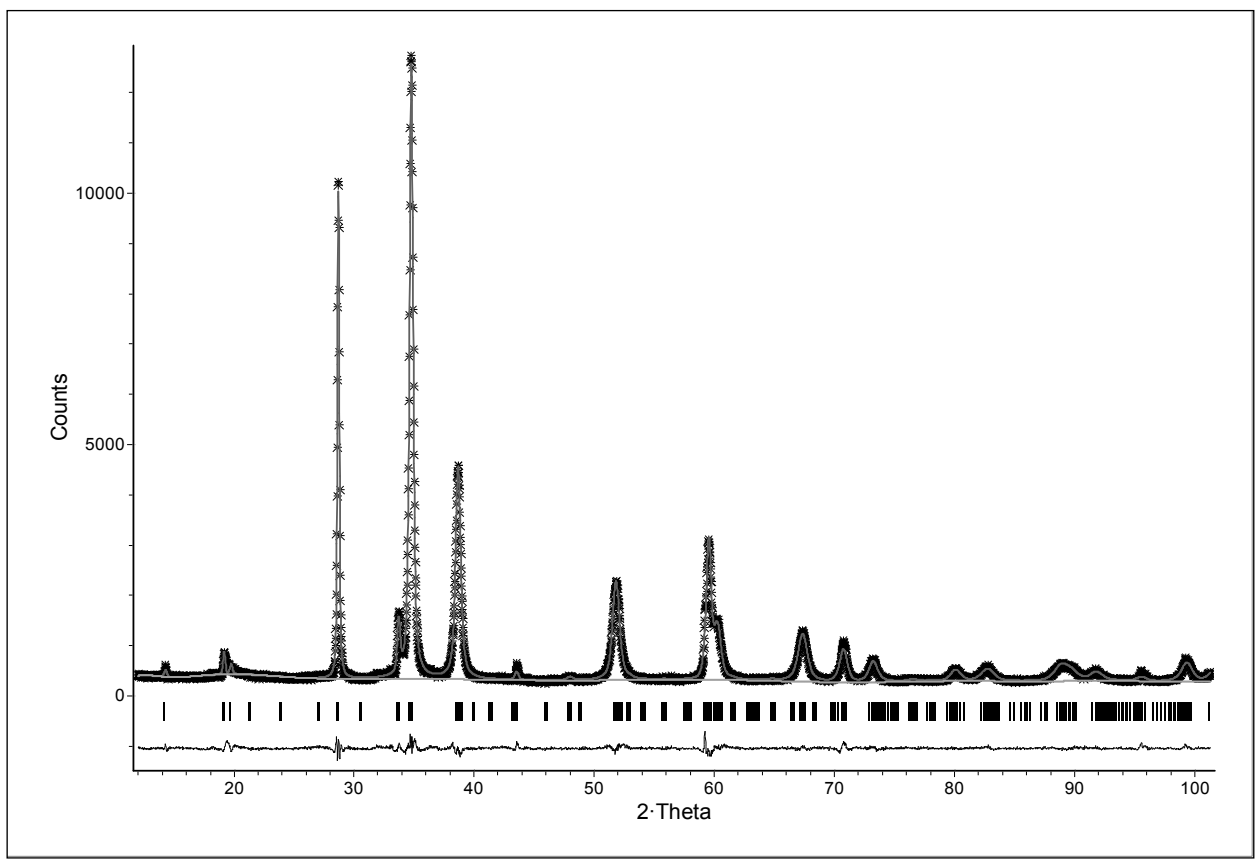

Fig. S2. XRD profile of $Ag_3Zn_2SbO_6$

Stars – experimental data, dark gray line – calculated pattern, light gray line – background, vertical bars – Bragg ticks, thin black line below – difference.

Table S2. Experimental and refinement details for $Ag_3M_2SbO_6$ (M = Co, Zn)

| Formula | $Ag_3Co_2SbO_6$ | $Ag_3Zn_2SbO_6$ |
|---|---|---|
| Formula weight | 659.2 | 672.1 |
| Crystal system | monoclinic | monoclinic |
| Space group | C2/m | C2/m |
| Z | 2 | 2 |
| a, Å | 5.3770(13) | 5.3829(2) |
| b, Å | 9.3118(22) | 9.3102(3) |
| c, Å | 6.4810(14) | 6.5046(2) |
| β, ° | 106.512(7) | 106.306(4) |
| V, Å$^3$ | 311.12(2) | 312.87(2) |
| Sample preparation | Amorphous admixture (beryllium carbonate or coffee) to reduce texture | |
| Diffractometer | Rigaku D/max-RC | ARL X'tra |
| Diffraction geometry | Bragg-Brentano | |
| Wavelength selection | Secondary-beam monochromator | Solid-state Si(Li) detector |
| Wavelength (CuKα), Å | 1.5406, 1.5444 | |
| U, kV | 55 | 40 |
| I, mA | 180 | 40 |
| Receiving slit, mm | 0.3 | 0.4 |
| Angular range, ° | 10–100 | 12–101.4 |
| Step size, ° | 0.02 | 0.02 |
| Count time, s | 3 | 2.4 |
| Number of data points | 4500 | 4470 |
| Number of hkl | 179 | 179 |
| Number of variables | 55 | 57 |
| $R_{wp}$ | 0.0769 | 0.0605 |
| $R_{exp}$ | 0.0276 | 0.0423 |
| $R_F^2$ | 0.0198 | 0.0228 |
| $\chi^2$ | 7.830 | 2.067 |